\documentclass[aps,prl,twocolumn,superscriptaddress]{revtex4-2}
\usepackage{xcolor,graphicx,amsmath}
\newcommand{\vecv}{\vec{V}}
\newcommand{\vel}{v}
\newcommand{\ud}{\mathrm{d}}
\newcommand{\rmd}{\mathrm{d}}
\newcommand{\dvx}{\Delta V_x}
\newcommand{\dvy}{\Delta V_y}
\newcommand{\ve}{\varepsilon}
\newcommand{\la}{\left\langle}
\newcommand{\ra}{\right\rangle}

\newcommand{\kB}{k_{\mathrm{B}}}

\newcommand*{\VEC}[1]{\boldsymbol{#1}}
\newcommand*{\TENSOR}[1]{{\mathsf{#1}}}

\DeclareMathOperator{\sx}{s}
\DeclareMathOperator{\cx}{c}

\begin{document}

\title{Sorting by Resetting}
\author{Bart Cleuren}
\email[]{bart.cleuren@uhasselt.be}
\affiliation{UHasselt, Faculty of Sciences, Theory Lab, Agoralaan, 3590 Diepenbeek, Belgium}
\author{Ralf Eichhorn}
\email[]{ralf.eichhorn@su.se}
\affiliation{Nordita, Royal Institute of Technology and Stockholm University, Hannes
Alfv\'ens v\"ag 12, SE-106 91 Stockholm, Sweden}

\date{\today}

\begin{abstract}
A novel paradigm for sorting is introduced, based upon resetting. Using simple examples, we demonstrate that sorting is achieved by resetting the velocity component(s) or orientation of the particles, rather than position. The objects to be sorted are microparticles, modeled as suspended and spatially extended Brownian particles. This sorting-by-resetting scheme illustrates that stochastic resetting can create non-equilibrium conditions which enable tasks forbidden at thermodynamic equilibrium.
\end{abstract}
\maketitle

In the context of stochastic processes, such as Brownian motion \cite{vanKampen2007Stochastic}, resetting refers to the mechanism in which the system’s natural evolution---its relaxation towards equilibrium---is intermittently interrupted, and the system’s configuration is returned to a prescribed state, from which the dynamics resumes as if freshly initiated. Resetting thus prevents the system from reaching its long-term behavior. Instead, it maintains a nontrivial non-equilibrium stationary state \cite{evansPRL2011,evansJPA2020}, sustained by the continuous cycle of relaxation and resetting. 
A hallmark application is the acceleration of search processes: resetting the position of the searcher effectively truncates unproductive excursions and optimizes first-passage times \cite{evansPRL2011,evansJPA2020}. In this Letter, we exploit the non-equilibrium nature of the relaxation-resetting cycle to perform tasks that are forbidden in thermodynamic equilibrium. Specifically, we focus on sorting microparticles based on their intrinsic properties, such as shape and mass.

The sorting of microparticles constitutes a crucial task in both academic and industrial contexts. These particles exhibit a wide range of sizes, shapes, and compositions \cite{championPNAS2007,wittman2023}, and such parameters critically influence their functional properties \cite{kinnear2017,csordas2019}. Consequently, the development of versatile and efficient techniques for their purification is of significant importance.  Prominent sorting mechanisms include structured microfluidic devices which exploit size- and shape-specific particle interactions with the topographical structure or the induced flow pattern to achieve separation \cite{regtmeier2007,bogunovicPRL2012,aristovSM2013,volpe2013,chen2015,sonkerARAC2019}, micro- or nanofluidic particle flows in which particles are separated by applying multiple external fields to the flow and/or the particles \cite{LEE2023114688}, and optical screening of individual particles to obtain shape-specific scattering signals which are then used for particle classification and sorting \cite{csordas2019}.

In the context of sorting, the conventional approach of resetting particles to a specific spatial location is evidently unsuitable, as it would merely lead to the accumulation and mixing of different particle types at that location. Requiring a uniform resetting protocol for all particle species—thus obviating the need for any form of pre-sorting—we propose an alternative strategy in which resetting acts on degrees of freedom other than position. Specifically, we reset velocities \cite{olsenJStat2024,santraCHAOS2025} or orientations, while particle positions are left unchanged.
\begin{figure}
\includegraphics[width=0.9\columnwidth]{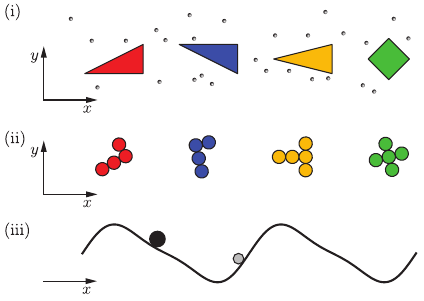}
\caption{Illustration of the three different scenarios.
(i) Tracer particles of various shapes with fixed orientation in an ideal gas.
(ii) Chiral and other colloidal particles suspended in an aqueous solution.
(iii) Underdamped spherical Brownian particles of different mass
(due to different sizes or densities)
in a one-dimensional asymmetric potential landscape.
}
\label{fig:models}
\end{figure}
We illustrate this \textit{sorting-by-resetting} principle in three different scenarios (see Fig.~\ref{fig:models}):
(i) Tracer particles of various (convex) shapes with fixed orientation in an ideal gas of point-like particles. Periodic resetting of their translational velocity results in net motion along shape-dependent directions. This setup is analyzed in two dimensions using kinetic theory.
(ii) A suspension of chiral and other colloidal particles. 
When their orientation is regularly reset to a common reference, particles of different shapes separate into distinct net directions of motion. This system is modeled by overdamped Langevin equations coupling position and orientation.
(iii) Underdamped spherical Brownian particles in a one-dimensional asymmetric potential landscape, described by Langevin equations for particle position and velocity. Repeatedly resetting the particle velocity induces a ratchet-like net displacement with mass-dependent average drift velocity. While spatial anisotropy in the first two setups is  ``intrinsic''  to the particles \cite{vandenBroek_2009} due to their various shapes, it here is provided ``externally'' by the potential landscape.

\begin{figure}[t!]
\includegraphics[width=0.85\columnwidth]{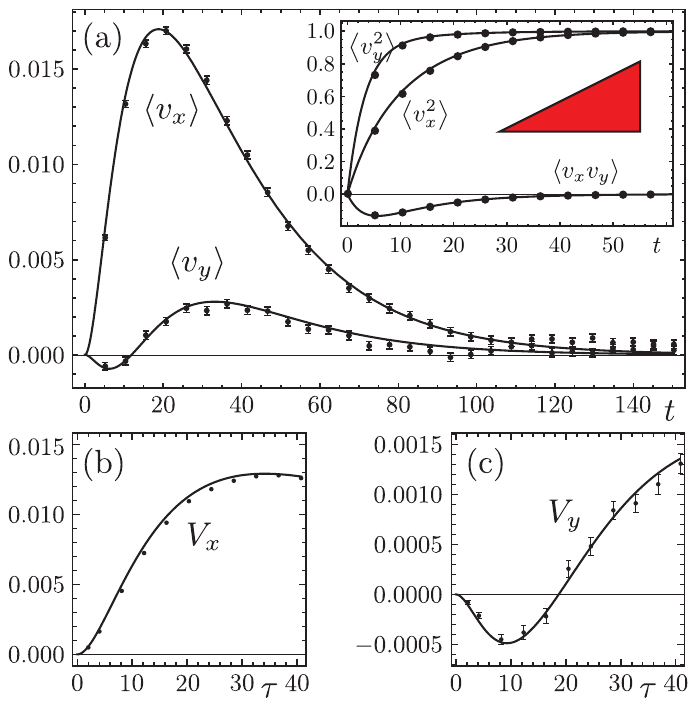}
\caption{(a) Relaxation of the lowest order moments for the rectangular shape object shown in the inset
($\ve = \sqrt{1/20} \approx 0.22$), in response to (re-)setting $\vel_x=0$ and $\vel_y=0$ at time $t=0$. The inset shows the relaxation of the second moments. Solid curves: theoretical predictions according to Eqs.~\eqref{eq:moments}. Dots: Simulation results, obtained as an average over $2 \times 10^7$ independent realizations (per data point). (b)-(c) Average displacement velocity of the same triangle as a function of the resetting period $\tau$. Solid lines: theoretical predictions according to Eq.~\eqref{eq:tracer:V}. Dots with standard error bars: simulation results obtained as an average over 500 realizations (per data point). Parameters: $S=1$, $\kB T=1$, $\rho=1$; the time unit is the mean free time between particle-object collisions, i.e.~the time values can also be read as the average number of collisions.
}
\label{fig:tracer_relaxation}
\end{figure}

\begin{figure}
\includegraphics[width=0.95\columnwidth]{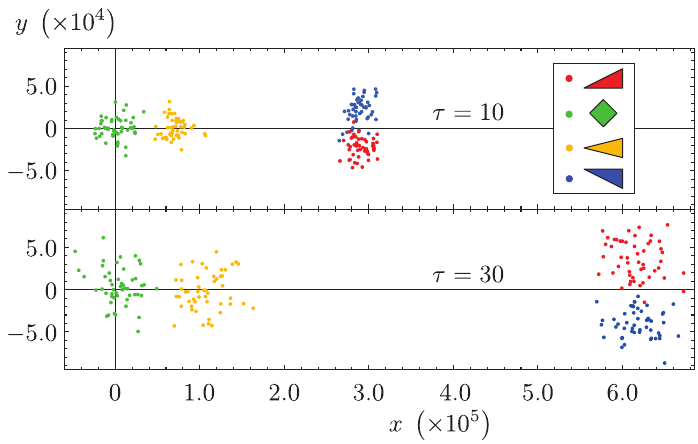}
\caption{Separation of four different kinds of tracer particles (see legend): shown are the positions of 50 particles
per tracer species after a time $5\times10^{7}$; all particles started at the origin at time 0.
Upper panel: resetting period $\tau = 10$. Lower panel: resetting period $\tau = 30$.
Parameters: $S=1$ (for all species), $\kB T=1$, $\rho=1$; the time unit is the mean free time between particle-object collisions.}
\label{fig:tracer_separation}
\end{figure}

\paragraph{(i) Tracer particle in an ideal gas.}
A spatially extended 2D object (mass $M$) experiences free movement between elastic collisions with the point-like particles (mass $m$) of a surrounding ideal gas. We focus solely on translational motion and keep the object orientation fixed at all times. The ideal gas is in equilibrium, at temperature $T$, so that, when left on its own, the object will equilibrate and perform undirected Brownian motion. In this regime, the velocity components are Gaussian distributed. Periodic resetting of the tracer object's velocity prevents it from equilibrating with the gas, but repeatedly restarts a relaxation process. Describing the translational motion of the tracer via the probability density of its velocity, a master equation can be formulated based on the elastic collisions with the gas particles that conserve energy and momentum. It is then possible to obtain an exact, closed, but infinite set of evolution equations for the velocity moments and their cross-correlations. Expanding this set of equations in $\varepsilon = \sqrt{m/M} \ll 1$ for heavy tracer objects
decouples them and leads to the following result for the first- and second-order moments (see the Appendix for more details; the full derivation is detailed in \cite{tracerTheory}, a similar approach has been used in \cite{cleurenJSTAT2023,wijns2023microscopic}):
\begin{subequations}
\label{eq:moments}
\begin{widetext}
\begin{eqnarray}
\frac{d}{d t}\langle \vel_x \rangle &=&
	4\ve^2 \Big[ \la\vel_y\ra \la\sx\cx\ra - \langle \vel_x \rangle \left\langle\sx^2  \right\rangle \Big]
    	-\sqrt{2\pi} \ve^3  \Big[  \la\vel_x^2\ra\la\sx^3\ra+\la\vel_y^2\ra\la\sx \cx^2\ra-2\la\vel_x\vel_y\ra\la\sx^2\cx\ra\Big]
\, , \\
\frac{d}{d t}\langle \vel_y \rangle &=&
	4\ve^2 \Big[ \la\vel_x\ra \la \sx\cx \ra - \la\vel_y \ra\la\cx^2\ra \Big]
	+\sqrt{2\pi}\ve^3  \Big[\la\vel_x^2\ra\la\sx^2\cx\ra+\la\vel_y^2\ra\la\cx^3\ra-2\la\vel_x\vel_y\ra\la\sx\cx^2\ra\Big]
\, ,
\end{eqnarray}
\end{widetext}
and
\begin{eqnarray}
\hspace{-4mm}&&\frac{d}{dt}\langle \vel_x^2 \rangle =
8\ve^2 \Big[ \la\sx^2\ra\big(1-\la\vel_x^2\ra\big)+\la\vel_x\vel_y\ra \la\sx\cx\ra \Big]
\;, \\
\hspace{-4mm}&&\frac{d}{dt}\langle \vel_x \vel_y \rangle =
4\ve^2 \Big[\big( \la\vel_x^2\ra+\la\vel_y^2\ra-2 \big) \la\sx\cx\ra -\la\vel_x\vel_y\ra\Big]
\;, \\
\hspace{-4mm}&&\frac{d}{dt}\langle \vel_y^2 \rangle =
8\ve^2 \Big[ \la\cx^2\ra\big(1-\la\vel_y^2\ra\big)+ \la\vel_x\vel_y\ra\la\sx\cx\ra \Big]
\;. 
\end{eqnarray}
\end{subequations}
The velocity is expressed in terms of the thermal velocity $\sqrt{\kB T/M}$, and the unit of time is the mean free time, $\bar{t}=\sqrt{2\pi m/\kB T}(S\rho)^{-1}$, defined as the average time between two consecutive collisions with the object. Here $\kB$ is Boltzmann's constant, $S$ the circumference of the object,
and $\rho$ the gas density. The shape of the tracer object is encoded in the goniometric averages along its boundary,
\begin{equation}\label{eq:gonio}
    \la\sx^n\cx^m\ra\equiv \int_{0}^{2\pi}\mathrm{d} \theta \, s(\theta)\sin^n(\theta) \cos^m(\theta) 
\, ,
\end{equation}
with $s(\theta)$ the shape function, defined such that $s(\theta)\mathrm{d} \theta$ is the fraction of the surface with orientation $\theta$ w.r.t the $x$-axis (see the Appendix). Different shapes have different coefficients appearing in the differential equations, and hence show different relaxation towards equilibrium. The resetting procedure corresponds to fixing the initial conditions for the Eqs.~\eqref{eq:moments}, which then evolve ``freely'' to the next resetting event. Figure \ref{fig:tracer_relaxation}(a) shows the relaxation of the lowest order velocity moments for a triangular tracer particle (with a right angle at its base), whose velocity has been reset to $v_x=0$, $v_y=0$ at $t=0$. The analytical solution from \eqref{eq:moments} shows excellent agreement with numerical simulations (details of the simulations are do\-cu\-mented in \cite{wijns2023microscopic}). Figures \ref{fig:tracer_relaxation}(b), (c) demonstrate that periodic resetting induces net motion of the tracer particle, with an average displacement velocity given by
\begin{equation}
\label{eq:tracer:V}
\VEC{V} =
\begin{pmatrix}
V_x
\\[0.7ex]
V_y
\end{pmatrix}
= \frac{1}{\tau}
\begin{pmatrix}
\int_0^\tau \mathrm{d}t \, \langle \vel_x \rangle
\\[1ex]
\int_0^\tau \mathrm{d}t \, \langle \vel_y \rangle
\end{pmatrix}
\, .
\end{equation}
Figure~\ref{fig:tracer_relaxation}(b) and (c) show excellent agreement between this theoretical
prediction and simulation results.
Separation of four different types of tracer particles is demonstrated in
Fig.~\ref{fig:tracer_separation}; all particles experience the same protocol of periodically resetting the particle velocity
to $\vel_x=0$ and $\vel_y=0$ at time-intervals $\tau$.

\begin{figure}
\includegraphics[width=0.85\columnwidth]{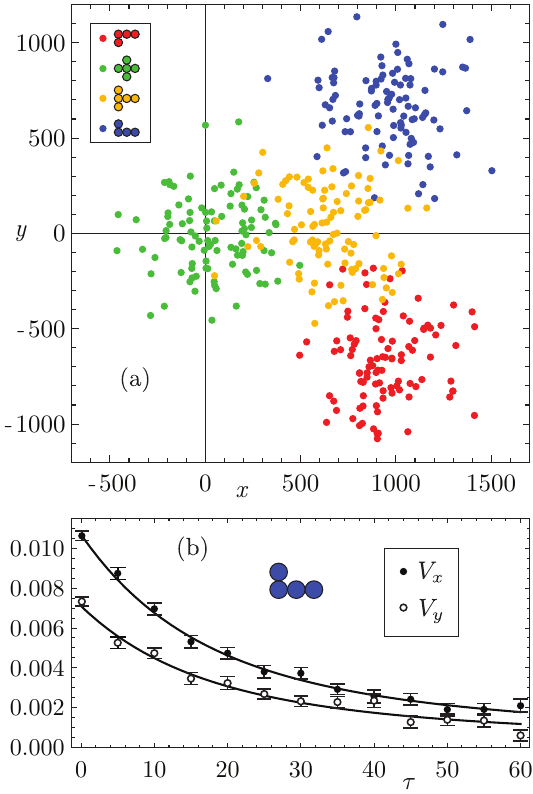}
\caption{(a) Separation of four differently shaped colloids (see legend). The dots are the positions of 100 colloids per species after time 90000. All particles start at the origin at time 0 and experience the same resetting protocol: orientation is reset along the $x$ axis (this is the orientation shown in the legends) in periodic intervals $\tau = 0.1$. The colloidal objects are assembled from beads of unit diameter as illustrated in the legend. Parameters: thermal energy $\kB T = 4$, viscosity of the solution $\nu = 1$ (this quantity enters the calculation of the friction tensors, see Appendix). When choosing water at room temperature as a solution for the colloidal particles,
these parameter values correspond to units of seconds for time scales and of micrometers for length scales. (b) The average displacement velocity (given in Eq.~\eqref{eq:colloidal:V}) for the shown particle as a function of the reset period $\tau$. }
\label{fig:colloid_separation}
\end{figure}

\paragraph{(ii) Suspension of colloidal particles.}
Non-spherical colloidal particles in suspension move and rotate by diffusion. A resetting event now affects particle orientation only, while its position is ``frozen''.
In two dimensions, the diffusive motion of a colloidal particle is modeled by the overdamped Langevin equations for position $x(t),y(t)$ and orientation $\varphi(t)$,
\begin{align}
\label{eq:colloids}
\dot{\VEC{q}}(t) &= \sqrt{2\kB T} \, \TENSOR{R}(\varphi) \TENSOR{\mu}^{1/2} \, \VEC{\xi}(t)
\, .
\end{align}
The coordinate $\VEC{q}(t)=(q_1(t),q_2(t),q_3(t))=(x(t),y(t),\varphi(t))$ collects all degrees of freedom,
$\VEC{\xi}(t)=(\xi_1(t),\xi_2(t),\xi_3(t))$ are mutually independent, unbiased, and $\delta$-correlated
white noise processes, and $T$ is the temperature of the thermal environment (aqueous solution).
The mobility tensor $\mu$ captures the hydrodynamic coupling between the three degrees of
freedom and is characteristic of the shape of the particle.
It is symmetric and positive definite, so its square root is well-defined, with $\mu^{1/2}\mu^{1/2}=\mu$. For the different particle species we consider in Fig.~\ref{fig:colloid_separation}, explicit expressions for $\mu$ in a \textit{body-fixed reference frame} are given in the Appendix, Eqs.~\eqref{eq:mus}.

The Langevin equations \eqref{eq:colloids} are written
in the \textit{laboratory system}, with the tensor
\begin{equation}
\TENSOR{R}(\varphi) = \left(
\begin{array}{ccc}
\cos\varphi & -\sin\varphi & 0 \\
\sin\varphi &  \cos\varphi & 0 \\
0 & 0 & 1 \\
\end{array}
\right)
\end{equation}
rotating from the body to the laboratory frame;
the multiplicative noise in \eqref{eq:colloids} is to be interpreted in the Stratonovich sense \cite{vanKampen2007Stochastic}.
From the  Langevin equations \eqref{eq:colloids} we can derive the evolution equations for the
moments of the particle position, using Ito's formula \cite{vanKampen2007Stochastic},
\begin{subequations}
\label{eq:<dx>and<dy>}
\begin{align}
\langle \dot{x}(t) \rangle &= \kB T \left[ -\mu_{13} \langle \sin\varphi(t) \rangle - \mu_{23} \langle \cos\varphi(t) \rangle \right]
\, , \\
\langle \dot{y}(t) \rangle &= \kB T \left[ +\mu_{13} \langle \cos\varphi(t) \rangle - \mu_{23} \langle \sin\varphi(t) \rangle \right]
\, ,
\end{align}
\end{subequations}
and for the moments $\langle \sin\varphi(t) \rangle$ and $\langle \cos\varphi(t) \rangle$.
The solutions of the latter equations are
$\langle \sin\varphi(t) \rangle = \langle \sin\varphi(0) \rangle \, e^{-\kB T \mu_{33} t}$
and
$\langle \cos\varphi(t) \rangle = \langle \cos\varphi(0) \rangle \, e^{-\kB T \mu_{33} t}$.
The quantity $\kB T \mu_{33}$ in the exponent is the rotational diffusion coefficient of the particle.

The evolution equations \eqref{eq:<dx>and<dy>} describe the particle displacements induced by rotation through
the hydrodynamic translation-rotation coupling, which is captured quantitatively in the coefficients $\mu_{13}$ and $\mu_{23}$
of the mobility tensor.
As expected, in the long term the average motion ceases as $\langle \sin\varphi(t) \rangle \xrightarrow{t \to \infty} 0$ and
$\langle \cos\varphi(t) \rangle \xrightarrow{t \to \infty} 0$.
However, net motion can be achieved by periodically reorienting the particles into a fixed
direction, and letting them diffuse freely between these resetting events, exploiting the transient dynamics of
\eqref{eq:<dx>and<dy>}. If we reset the particles to an angle $\varphi_0$, the transient motion is
\begin{subequations}
\label{eq:<x>and<y>}
\begin{align}
\langle x(t) \rangle &= x_0
- \left( \frac{\mu_{13}}{\mu_{33}} \sin\varphi_0 + \frac{\mu_{23}}{\mu_{33}} \cos\varphi_0 \right)
	\left( 1- e^{-\kB T \mu_{33} t} \right)
\, , \\
\langle y(t) \rangle &= y_0
+ \left( \frac{\mu_{13}}{\mu_{33}} \cos\varphi_0 - \frac{\mu_{23}}{\mu_{33}} \sin\varphi_0 \right)
	\left( 1- e^{-\kB T \mu_{33} t} \right)
\, ,
\end{align}
\end{subequations}
with $(x_0,y_0)$ being the initial particle position at the resetting event.
We recall that the components of the mobility tensor are characteristic of the shape of the particle so, in general, different particle species will perform different transient motions, leading to their spatial separation. Moreover, the direction of the transient motion can be controlled by the resetting angle $\varphi_0$.

In Figure \ref{fig:colloid_separation}, we illustrate this separation mechanism. The average particle current resulting from resetting to an orientation $\varphi_0=0$ at time-intervals $\tau$ is directly obtained from \eqref{eq:<x>and<y>},
\begin{equation}
\label{eq:colloidal:V}
\VEC{V} = \frac{1}{\tau}
\begin{pmatrix} -\frac{\mu_{23}}{\mu_{33}} \\[1ex] \frac{\mu_{13}}{\mu_{33}} \end{pmatrix}
\left( 1- e^{-\kB T \mu_{33} \tau} \right)
\, .
\end{equation}
Interestingly, this result predicts a monotonically decreasing particle velocity as a function of the resetting period $\tau$, with its maximum value $(-\kB \mu_{23}, \kB \mu_{13})$ for infinitely fast resetting $\tau \to 0$ and approaching $(0,0)$ as $\tau \to \infty$ (because then the system is effectively in equilibrium).

\paragraph{(iii) Underdamped Brownian ratchet.}
A massive Brownian particle (mass $m$) moving in a one-dimensional ratchet potential  \cite{reimannPR2002}
of spatial period $L$ and characteristic energy scale $U_0$,
\begin{equation}
U(x) = U_0 \left[ \sin( 2\pi x/L ) + 0.25 \sin( 4\pi x/L ) \right]
\, ,
\end{equation}
is described by the Langevin equation for position $x$ and velocity $v$ \cite{vanKampen2007Stochastic}
\begin{subequations}
\begin{align}
\dot{x}(t) &= v(t)
\, , \\
\dot{v}(t) &= -\frac{1}{m} U'(x(t)) - \frac{\gamma}{m} v(t) + \frac{1}{m} \sqrt{2 \kB T \gamma} \, \xi(t)
\, .
\end{align}
\end{subequations}
Here, $\gamma$ is the friction coefficient of the particle, $T$ the temperature of the thermal bath,
and $\xi(t)$ is an unbiased, $\delta$-correlated Gaussian white noise process.

\begin{figure}
\includegraphics[width=0.85\columnwidth]{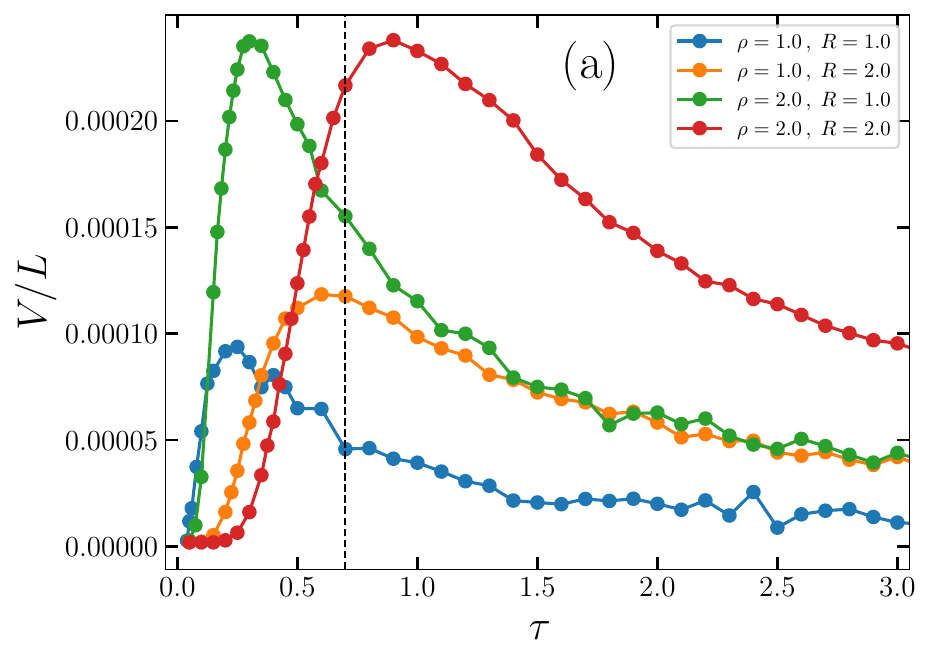}
\\
\includegraphics[width=0.85\columnwidth]{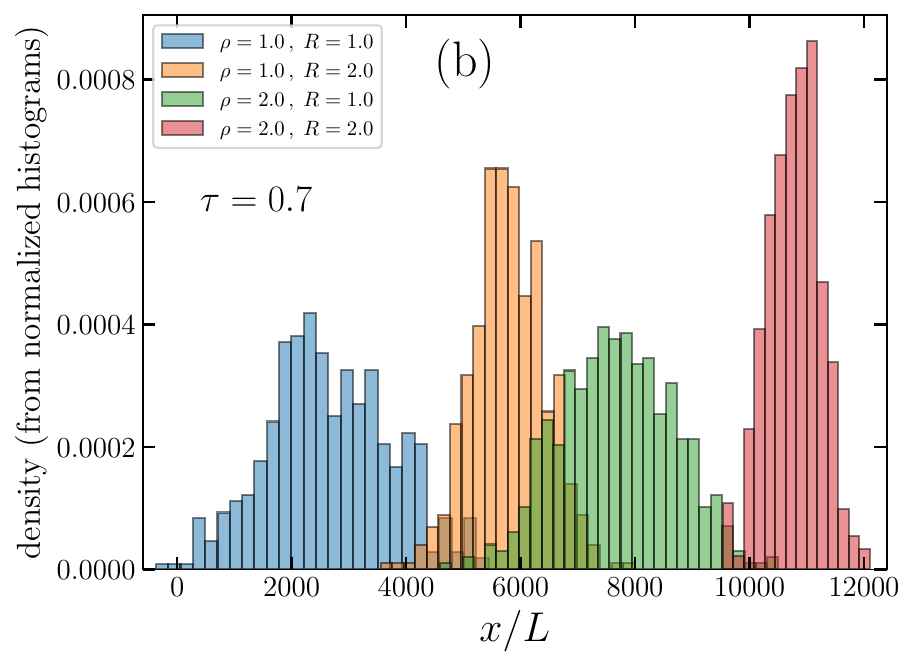}
\caption{Separation of spherical particles in a spatially asymmetric potential.
(a) Average velocity as a function of $\tau$ for four different particle species, combining two different particle densities $\rho$ with two different particle radii $R$ (see legend). The dots are numerical data obtained from averaging over 20000 independent realizations per data point, lines are a guide to the eye, error bars (not shown) are about the symbol size;
the spatial component is given in units of the period length $L$.
(b) Histograms of the distribution of 500 particles per species after running the resetting protocol with $\tau=0.7$ (dashed line in (a)) for
a total time of $5\times10^7$.
Other parameters: $U_0=8$, $L=0.5$, $\kB T=4$; the particle mass is $m=(4\pi/3) \rho R^3$, and the friction coefficient $\gamma=6\pi\nu R$ with
the viscosity $\nu=10$.
Translating these dimensionless parameters into dimensional quantities, one obtains length scales of $\mu$m, time scales of ms, $\kB T$=4 corresponds to the thermal energy at room temperature, $\rho=1$ corresponds to the density of water, and $\nu=10$ corresponds to 1\% of the viscosity of water.}
\label{fig:ratchet_separation}
\end{figure}

When left to itself, the particle will slowly diffuse along the ratchet potential with some transient, potentially directional
dynamics depending on its initial conditions $(x_0,v_0)$, but without preferential direction of motion in the long run, i.e.\ its average, long-term velocity
$V = \lim_{t \to \infty} \langle x(t) \rangle/t$
will be zero.
However, regularly resetting the particle's intrinsic velocity $v$ to a fixed value introduces non-equilibrium conditions,
which ``exploit'' the transient behavior to generate directional motion through a ratchet-like mechanism.
Due to the nonlinearity of the potential $U(x)$, a simple theoretical description, like in the previous two examples, is not available. However, the numerical simulations shown in Fig.\ \ref{fig:ratchet_separation} confirm our intuitive expectations,
and furthermore demonstrate that the net direction of motion depends on particle properties, e.g.\ its mass.
Velocity resetting thus sorts different particle species moving in the same ratchet potential.

To conclude, we introduced a novel paradigm for sorting colloidal particles according to their shape, size, or other characteristics (e.g.~mass). The sorting procedure relies on periodically resetting non-positional degrees of freedom, such as velocity and orientation. Each reset drives the particles transiently out of equilibrium with their surroundings. Under spatially asymmetric conditions---arising either from particle properties or from the environment---the subsequent relaxation towards equilibrium becomes biased, resulting in a net displacement. Repeated resetting establishes a non-equilibrium steady state and produces a systematic drift whose direction and magnitude depend on the particle characteristics, as well as the resetting frequency.

As theoretical proof-of-concept for this sorting scheme we considered three conceptually different scenarios (see Fig.~\ref{fig:models}): non-trivially shaped tracer particles in a dilute ideal gas (cf.~Fig.~\ref{fig:tracer_separation}), colloidal particles of various shapes suspended in an aqueous solution (overdamped regime, cf.~Fig.~\ref{fig:colloid_separation}), and underdamped Brownian particles in an asymmetric potential (cf.~Fig.~\ref{fig:ratchet_separation}). In all three scenarios, an outspoken sorting effect is observed, illustrating the robustness of the sorting scheme. In the first two scenarios, sorting stems solely from the characteristics of the objects to be sorted, and the resetting. In the third scenario, with spherical particles, the external potential is necessary to break spatial symmetry. Apart from that, there is no need for any purpose-built environment. This feature makes our method highly flexible.

From an experimental or technological viewpoint, resetting particle velocities in a controlled way appears challenging. For particles moving on a surface, switchable particle–surface interactions (e.g., via light-controlled surface chemistry \cite{KlajnCSR2010,LiuACIE2012,LiuCR2026}) can be used to immobilize particles by pinning them to the surface. Resetting orientational degrees of freedom can be achieved with electric fields, if the particles possess a dipole moment or are polarizable (note that the position of the particles needs to be kept fixed during resetting, e.g., again by pinning them to a surface).

Evidently, in our three scenarios, there are alternative options for sorting. For instance, tracer particles with fixed orientation can also be separated by applying a constant external force (likewise for the overdamped colloidal particles in solution, provided their orientation is pinned simultaneously). However, under the resetting procedure, not all particle species perform a net displacement, so it can be used to purify particle mixtures by removing undesired contaminants. Since the resetting protocol affects the second moments as well as particle velocities and displacements, sorting might even be achieved by selectively enhancing the diffusion of a specific particle species. As a complementary application, we envision diffusive mixing (in particular at low Reynolds numbers), induced by a resetting protocol that enhances the diffusion of \emph{all} particles.

We here presented only the most basic setup of the sorting-by-resetting scheme: periodic resetting to a fixed velocity or orientation. All the different aspects discussed above can be studied with more flexible or elaborate protocols, e.g., random resetting \cite{evansPRL2011,evansJPA2020} or resetting to distributions (different from the equilibrium distribution of the relevant degree of freedom) rather than fixed values \cite{evansJPA2020}. Moreover, we hope our work inspires further exploration of resetting for tasks other than sorting that require non-equilibrium conditions. The key point of our sorting-by-resetting scheme is to demonstrate that resetting specific degrees of freedom, different from the customary spatial positions, can lead to novel phenomena and applications.

\begin{acknowledgments}
R.E. acknowledges funding by the Swedish Research Council (Vetenskapsrådet) under Grants No.~2024-05091 and No.~638-2013-9243. B.C. would like to thank Nordita for the hospitality and acknowledges financial support from the Research Foundation - Flanders (FWO) under Grant No. V447425N. The resources and services used in this work were partly provided by the VSC (Flemish Supercomputer Center), funded by the Research Foundation - Flanders (FWO) and the Flemish Government.
\end{acknowledgments}
\bibliography{sores.bib}

\begin{thebibliography}{27}%
\makeatletter
\providecommand \@ifxundefined [1]{%
 \@ifx{#1\undefined}
}%
\providecommand \@ifnum [1]{%
 \ifnum #1\expandafter \@firstoftwo
 \else \expandafter \@secondoftwo
 \fi
}%
\providecommand \@ifx [1]{%
 \ifx #1\expandafter \@firstoftwo
 \else \expandafter \@secondoftwo
 \fi
}%
\providecommand \natexlab [1]{#1}%
\providecommand \enquote  [1]{``#1''}%
\providecommand \bibnamefont  [1]{#1}%
\providecommand \bibfnamefont [1]{#1}%
\providecommand \citenamefont [1]{#1}%
\providecommand \href@noop [0]{\@secondoftwo}%
\providecommand \href [0]{\begingroup \@sanitize@url \@href}%
\providecommand \@href[1]{\@@startlink{#1}\@@href}%
\providecommand \@@href[1]{\endgroup#1\@@endlink}%
\providecommand \@sanitize@url [0]{\catcode `\\12\catcode `\$12\catcode
  `\&12\catcode `\#12\catcode `\^12\catcode `\_12\catcode `\%12\relax}%
\providecommand \@@startlink[1]{}%
\providecommand \@@endlink[0]{}%
\providecommand \url  [0]{\begingroup\@sanitize@url \@url }%
\providecommand \@url [1]{\endgroup\@href {#1}{\urlprefix }}%
\providecommand \urlprefix  [0]{URL }%
\providecommand \Eprint [0]{\href }%
\providecommand \doibase [0]{https://doi.org/}%
\providecommand \selectlanguage [0]{\@gobble}%
\providecommand \bibinfo  [0]{\@secondoftwo}%
\providecommand \bibfield  [0]{\@secondoftwo}%
\providecommand \translation [1]{[#1]}%
\providecommand \BibitemOpen [0]{}%
\providecommand \bibitemStop [0]{}%
\providecommand \bibitemNoStop [0]{.\EOS\space}%
\providecommand \EOS [0]{\spacefactor3000\relax}%
\providecommand \BibitemShut  [1]{\csname bibitem#1\endcsname}%
\let\auto@bib@innerbib\@empty
\bibitem [{\citenamefont {Van~Kampen}(2007)}]{vanKampen2007Stochastic}%
  \BibitemOpen
  \bibfield  {author} {\bibinfo {author} {\bibfnamefont {N.~G.}\ \bibnamefont
  {Van~Kampen}},\ }\href@noop {} {\emph {\bibinfo {title} {Stochastic processes
  in physics and chemistry}}},\ \bibinfo {edition} {3rd}\ ed.\ (\bibinfo
  {publisher} {Elsevier},\ \bibinfo {address} {Amsterdam},\ \bibinfo {year}
  {2007})\BibitemShut {NoStop}%
\bibitem [{\citenamefont {Evans}\ and\ \citenamefont
  {Majumdar}(2011)}]{evansPRL2011}%
  \BibitemOpen
  \bibfield  {author} {\bibinfo {author} {\bibfnamefont {M.~R.}\ \bibnamefont
  {Evans}}\ and\ \bibinfo {author} {\bibfnamefont {S.~N.}\ \bibnamefont
  {Majumdar}},\ }\bibfield  {title} {\bibinfo {title} {Diffusion with
  stochastic resetting},\ }\href@noop {} {\bibfield  {journal} {\bibinfo
  {journal} {Phys. Rev. Lett.}\ }\textbf {\bibinfo {volume} {106}},\ \bibinfo
  {pages} {160601} (\bibinfo {year} {2011})}\BibitemShut {NoStop}%
\bibitem [{\citenamefont {Evans}\ \emph {et~al.}(2020)\citenamefont {Evans},
  \citenamefont {Majumdar},\ and\ \citenamefont {Schehr}}]{evansJPA2020}%
  \BibitemOpen
  \bibfield  {author} {\bibinfo {author} {\bibfnamefont {M.~R.}\ \bibnamefont
  {Evans}}, \bibinfo {author} {\bibfnamefont {S.~N.}\ \bibnamefont
  {Majumdar}},\ and\ \bibinfo {author} {\bibfnamefont {G.}~\bibnamefont
  {Schehr}},\ }\bibfield  {title} {\bibinfo {title} {Stochastic resetting and
  applications},\ }\href@noop {} {\bibfield  {journal} {\bibinfo  {journal} {J.
  Phys. A: Math. Theor.}\ }\textbf {\bibinfo {volume} {53}},\ \bibinfo {pages}
  {193001} (\bibinfo {year} {2020})}\BibitemShut {NoStop}%
\bibitem [{\citenamefont {Champion}\ \emph {et~al.}(2007)\citenamefont
  {Champion}, \citenamefont {Katare},\ and\ \citenamefont
  {Mitragotri}}]{championPNAS2007}%
  \BibitemOpen
  \bibfield  {author} {\bibinfo {author} {\bibfnamefont {J.~A.}\ \bibnamefont
  {Champion}}, \bibinfo {author} {\bibfnamefont {Y.~K.}\ \bibnamefont
  {Katare}},\ and\ \bibinfo {author} {\bibfnamefont {S.}~\bibnamefont
  {Mitragotri}},\ }\bibfield  {title} {\bibinfo {title} {Making polymeric
  micro- and nanoparticles of complex shapes},\ }\href@noop {} {\bibfield
  {journal} {\bibinfo  {journal} {PNAS}\ }\textbf {\bibinfo {volume} {104}},\
  \bibinfo {pages} {11901} (\bibinfo {year} {2007})}\BibitemShut {NoStop}%
\bibitem [{\citenamefont {Wittmann}\ \emph {et~al.}(2023)\citenamefont
  {Wittmann}, \citenamefont {Henze}, \citenamefont {Yan}, \citenamefont
  {Sharma},\ and\ \citenamefont {Simmchen}}]{wittman2023}%
  \BibitemOpen
  \bibfield  {author} {\bibinfo {author} {\bibfnamefont {M.}~\bibnamefont
  {Wittmann}}, \bibinfo {author} {\bibfnamefont {K.}~\bibnamefont {Henze}},
  \bibinfo {author} {\bibfnamefont {K.}~\bibnamefont {Yan}}, \bibinfo {author}
  {\bibfnamefont {V.}~\bibnamefont {Sharma}},\ and\ \bibinfo {author}
  {\bibfnamefont {J.}~\bibnamefont {Simmchen}},\ }\bibfield  {title} {\bibinfo
  {title} {Rod-shaped microparticles - an overview of synthesis and
  properties},\ }\href@noop {} {\bibfield  {journal} {\bibinfo  {journal}
  {Colloid and Polymer Science}\ }\textbf {\bibinfo {volume} {301}},\ \bibinfo
  {pages} {783} (\bibinfo {year} {2023})}\BibitemShut {NoStop}%
\bibitem [{\citenamefont {Kinnear}\ \emph {et~al.}(2017)\citenamefont
  {Kinnear}, \citenamefont {Moore}, \citenamefont {Rodriguez-Lorenzo},
  \citenamefont {Rothen-Rutishauser},\ and\ \citenamefont
  {Petri-Fink}}]{kinnear2017}%
  \BibitemOpen
  \bibfield  {author} {\bibinfo {author} {\bibfnamefont {C.}~\bibnamefont
  {Kinnear}}, \bibinfo {author} {\bibfnamefont {T.~L.}\ \bibnamefont {Moore}},
  \bibinfo {author} {\bibfnamefont {L.}~\bibnamefont {Rodriguez-Lorenzo}},
  \bibinfo {author} {\bibfnamefont {B.}~\bibnamefont {Rothen-Rutishauser}},\
  and\ \bibinfo {author} {\bibfnamefont {A.}~\bibnamefont {Petri-Fink}},\
  }\bibfield  {title} {\bibinfo {title} {Form follows function: Nanoparticle
  shape and its implications for nanomedicine},\ }\href@noop {} {\bibfield
  {journal} {\bibinfo  {journal} {Chemical Reviews}\ }\textbf {\bibinfo
  {volume} {117}},\ \bibinfo {pages} {11476} (\bibinfo {year}
  {2017})}\BibitemShut {NoStop}%
\bibitem [{\citenamefont {Mage}\ \emph {et~al.}(2019)\citenamefont {Mage},
  \citenamefont {Csordas}, \citenamefont {Brown}, \citenamefont {Klinger},
  \citenamefont {Eisenstein}, \citenamefont {Mitragotri}, \citenamefont
  {Hawker},\ and\ \citenamefont {Soh}}]{csordas2019}%
  \BibitemOpen
  \bibfield  {author} {\bibinfo {author} {\bibfnamefont {P.~L.}\ \bibnamefont
  {Mage}}, \bibinfo {author} {\bibfnamefont {A.~T.}\ \bibnamefont {Csordas}},
  \bibinfo {author} {\bibfnamefont {T.}~\bibnamefont {Brown}}, \bibinfo
  {author} {\bibfnamefont {D.}~\bibnamefont {Klinger}}, \bibinfo {author}
  {\bibfnamefont {M.}~\bibnamefont {Eisenstein}}, \bibinfo {author}
  {\bibfnamefont {S.}~\bibnamefont {Mitragotri}}, \bibinfo {author}
  {\bibfnamefont {C.}~\bibnamefont {Hawker}},\ and\ \bibinfo {author}
  {\bibfnamefont {H.~T.}\ \bibnamefont {Soh}},\ }\bibfield  {title} {\bibinfo
  {title} {Shape-based separation of synthetic microparticles},\ }\href@noop {}
  {\bibfield  {journal} {\bibinfo  {journal} {Nature Materials}\ }\textbf
  {\bibinfo {volume} {18}},\ \bibinfo {pages} {82} (\bibinfo {year}
  {2019})}\BibitemShut {NoStop}%
\bibitem [{\citenamefont {Regtmeier}\ \emph {et~al.}(2007)\citenamefont
  {Regtmeier}, \citenamefont {Eichhorn}, \citenamefont {Duong}, \citenamefont
  {Reimann}, \citenamefont {Anselmetti},\ and\ \citenamefont
  {Ros}}]{regtmeier2007}%
  \BibitemOpen
  \bibfield  {author} {\bibinfo {author} {\bibfnamefont {J.}~\bibnamefont
  {Regtmeier}}, \bibinfo {author} {\bibfnamefont {R.}~\bibnamefont {Eichhorn}},
  \bibinfo {author} {\bibfnamefont {T.~T.}\ \bibnamefont {Duong}}, \bibinfo
  {author} {\bibfnamefont {P.}~\bibnamefont {Reimann}}, \bibinfo {author}
  {\bibfnamefont {D.}~\bibnamefont {Anselmetti}},\ and\ \bibinfo {author}
  {\bibfnamefont {A.}~\bibnamefont {Ros}},\ }\bibfield  {title} {\bibinfo
  {title} {Pulsed-field separation of particles in a microfluidic device},\
  }\href@noop {} {\bibfield  {journal} {\bibinfo  {journal} {Eur. Phys. J. E}\
  }\textbf {\bibinfo {volume} {22}},\ \bibinfo {pages} {335} (\bibinfo {year}
  {2007})}\BibitemShut {NoStop}%
\bibitem [{\citenamefont {Bogunovic}\ \emph {et~al.}(2012)\citenamefont
  {Bogunovic}, \citenamefont {Fliedner}, \citenamefont {Eichhorn},
  \citenamefont {Wegener}, \citenamefont {Regtmeier}, \citenamefont
  {Anselmetti},\ and\ \citenamefont {Reimann}}]{bogunovicPRL2012}%
  \BibitemOpen
  \bibfield  {author} {\bibinfo {author} {\bibfnamefont {L.}~\bibnamefont
  {Bogunovic}}, \bibinfo {author} {\bibfnamefont {M.}~\bibnamefont {Fliedner}},
  \bibinfo {author} {\bibfnamefont {R.}~\bibnamefont {Eichhorn}}, \bibinfo
  {author} {\bibfnamefont {S.}~\bibnamefont {Wegener}}, \bibinfo {author}
  {\bibfnamefont {J.}~\bibnamefont {Regtmeier}}, \bibinfo {author}
  {\bibfnamefont {D.}~\bibnamefont {Anselmetti}},\ and\ \bibinfo {author}
  {\bibfnamefont {P.}~\bibnamefont {Reimann}},\ }\bibfield  {title} {\bibinfo
  {title} {Chiral particle separation by a nonchiral microlattice},\
  }\href@noop {} {\bibfield  {journal} {\bibinfo  {journal} {Phys. Rev. Lett.}\
  }\textbf {\bibinfo {volume} {109}},\ \bibinfo {pages} {100603} (\bibinfo
  {year} {2012})}\BibitemShut {NoStop}%
\bibitem [{\citenamefont {Aristov}\ \emph {et~al.}(2013)\citenamefont
  {Aristov}, \citenamefont {Eichhorn},\ and\ \citenamefont
  {Bechinger}}]{aristovSM2013}%
  \BibitemOpen
  \bibfield  {author} {\bibinfo {author} {\bibfnamefont {M.}~\bibnamefont
  {Aristov}}, \bibinfo {author} {\bibfnamefont {R.}~\bibnamefont {Eichhorn}},\
  and\ \bibinfo {author} {\bibfnamefont {C.}~\bibnamefont {Bechinger}},\
  }\bibfield  {title} {\bibinfo {title} {Separation of chiral colloidal
  particles in a helical flow field},\ }\href@noop {} {\bibfield  {journal}
  {\bibinfo  {journal} {Soft Matter}\ }\textbf {\bibinfo {volume} {9}},\
  \bibinfo {pages} {2525} (\bibinfo {year} {2013})}\BibitemShut {NoStop}%
\bibitem [{\citenamefont {Mijalkov}\ and\ \citenamefont
  {Volpe}(2013)}]{volpe2013}%
  \BibitemOpen
  \bibfield  {author} {\bibinfo {author} {\bibfnamefont {M.}~\bibnamefont
  {Mijalkov}}\ and\ \bibinfo {author} {\bibfnamefont {G.}~\bibnamefont
  {Volpe}},\ }\bibfield  {title} {\bibinfo {title} {Sorting of chiral
  microswimmers},\ }\href@noop {} {\bibfield  {journal} {\bibinfo  {journal}
  {Soft Matter}\ }\textbf {\bibinfo {volume} {9}},\ \bibinfo {pages} {6376}
  (\bibinfo {year} {2013})}\BibitemShut {NoStop}%
\bibitem [{\citenamefont {Chen}\ and\ \citenamefont {Ai}(2015)}]{chen2015}%
  \BibitemOpen
  \bibfield  {author} {\bibinfo {author} {\bibfnamefont {Q.}~\bibnamefont
  {Chen}}\ and\ \bibinfo {author} {\bibfnamefont {B.-q.}\ \bibnamefont {Ai}},\
  }\bibfield  {title} {\bibinfo {title} {{Sorting of chiral active particles
  driven by rotary obstacles}},\ }\href@noop {} {\bibfield  {journal} {\bibinfo
   {journal} {The Journal of Chemical Physics}\ }\textbf {\bibinfo {volume}
  {143}},\ \bibinfo {pages} {104113} (\bibinfo {year} {2015})}\BibitemShut
  {NoStop}%
\bibitem [{\citenamefont {Sonker}\ \emph {et~al.}(2019)\citenamefont {Sonker},
  \citenamefont {Kim}, \citenamefont {Egatz-Gomez},\ and\ \citenamefont
  {Ros}}]{sonkerARAC2019}%
  \BibitemOpen
  \bibfield  {author} {\bibinfo {author} {\bibfnamefont {M.}~\bibnamefont
  {Sonker}}, \bibinfo {author} {\bibfnamefont {D.}~\bibnamefont {Kim}},
  \bibinfo {author} {\bibfnamefont {A.}~\bibnamefont {Egatz-Gomez}},\ and\
  \bibinfo {author} {\bibfnamefont {A.}~\bibnamefont {Ros}},\ }\bibfield
  {title} {\bibinfo {title} {Separation phenomena in tailored micro-and
  nanofluidic environments},\ }\href@noop {} {\bibfield  {journal} {\bibinfo
  {journal} {Annual Review of Analytical Chemistry}\ }\textbf {\bibinfo
  {volume} {12}},\ \bibinfo {pages} {475} (\bibinfo {year} {2019})}\BibitemShut
  {NoStop}%
\bibitem [{\citenamefont {Lee}\ \emph {et~al.}(2023)\citenamefont {Lee},
  \citenamefont {Mishra},\ and\ \citenamefont {Kim}}]{LEE2023114688}%
  \BibitemOpen
  \bibfield  {author} {\bibinfo {author} {\bibfnamefont {K.}~\bibnamefont
  {Lee}}, \bibinfo {author} {\bibfnamefont {R.}~\bibnamefont {Mishra}},\ and\
  \bibinfo {author} {\bibfnamefont {T.}~\bibnamefont {Kim}},\ }\bibfield
  {title} {\bibinfo {title} {Review of micro/nanofluidic particle separation
  mechanisms: Toward combined multiple physical fields for nanoparticles},\
  }\href@noop {} {\bibfield  {journal} {\bibinfo  {journal} {Sensors and
  Actuators A: Physical}\ }\textbf {\bibinfo {volume} {363}},\ \bibinfo {pages}
  {114688} (\bibinfo {year} {2023})}\BibitemShut {NoStop}%
\bibitem [{\citenamefont {Olsen}\ and\ \citenamefont
  {L{\"o}wen}(2024)}]{olsenJStat2024}%
  \BibitemOpen
  \bibfield  {author} {\bibinfo {author} {\bibfnamefont {K.~S.}\ \bibnamefont
  {Olsen}}\ and\ \bibinfo {author} {\bibfnamefont {H.}~\bibnamefont
  {L{\"o}wen}},\ }\bibfield  {title} {\bibinfo {title} {Dynamics of inertial
  particles under velocity resetting},\ }\href@noop {} {\bibfield  {journal}
  {\bibinfo  {journal} {Journal of Statistical Mechanics: Theory and
  Experiment}\ }\textbf {\bibinfo {volume} {2024}},\ \bibinfo {pages} {033210}
  (\bibinfo {year} {2024})}\BibitemShut {NoStop}%
\bibitem [{\citenamefont {Santra}\ and\ \citenamefont
  {St{\o}levik~Olsen}(2025)}]{santraCHAOS2025}%
  \BibitemOpen
  \bibfield  {author} {\bibinfo {author} {\bibfnamefont {I.}~\bibnamefont
  {Santra}}\ and\ \bibinfo {author} {\bibfnamefont {K.}~\bibnamefont
  {St{\o}levik~Olsen}},\ }\bibfield  {title} {\bibinfo {title} {Brownian motion
  with stochastic energy renewals},\ }\href@noop {} {\bibfield  {journal}
  {\bibinfo  {journal} {Chaos: An Interdisciplinary Journal of Nonlinear
  Science}\ }\textbf {\bibinfo {volume} {35}} (\bibinfo {year}
  {2025})}\BibitemShut {NoStop}%
\bibitem [{\citenamefont {van~den Broek}\ \emph {et~al.}(2009)\citenamefont
  {van~den Broek}, \citenamefont {Eichhorn},\ and\ \citenamefont {Van~den
  Broeck}}]{vandenBroek_2009}%
  \BibitemOpen
  \bibfield  {author} {\bibinfo {author} {\bibfnamefont {M.}~\bibnamefont
  {van~den Broek}}, \bibinfo {author} {\bibfnamefont {R.}~\bibnamefont
  {Eichhorn}},\ and\ \bibinfo {author} {\bibfnamefont {C.}~\bibnamefont
  {Van~den Broeck}},\ }\bibfield  {title} {\bibinfo {title} {Intrinsic
  ratchets},\ }\href@noop {} {\bibfield  {journal} {\bibinfo  {journal}
  {Europhys. Lett.}\ }\textbf {\bibinfo {volume} {86}},\ \bibinfo {pages}
  {30002} (\bibinfo {year} {2009})}\BibitemShut {NoStop}%
\bibitem [{\citenamefont {Cleuren}\ and\ \citenamefont
  {Eichhorn}(2026)}]{tracerTheory}%
  \BibitemOpen
  \bibfield  {author} {\bibinfo {author} {\bibfnamefont {B.}~\bibnamefont
  {Cleuren}}\ and\ \bibinfo {author} {\bibfnamefont {R.}~\bibnamefont
  {Eichhorn}},\ }\href@noop {} {\bibinfo {title} {Tbd}},\ \bibinfo
  {howpublished} {in preparation} (\bibinfo {year} {2026})\BibitemShut
  {NoStop}%
\bibitem [{\citenamefont {Cleuren}\ and\ \citenamefont
  {Eichhorn}(2023)}]{cleurenJSTAT2023}%
  \BibitemOpen
  \bibfield  {author} {\bibinfo {author} {\bibfnamefont {B.}~\bibnamefont
  {Cleuren}}\ and\ \bibinfo {author} {\bibfnamefont {R.}~\bibnamefont
  {Eichhorn}},\ }\bibfield  {title} {\bibinfo {title} {Energetics of a
  microscopic feynman ratchet},\ }\href@noop {} {\bibfield  {journal} {\bibinfo
   {journal} {Journal of Statistical Mechanics: Theory and Experiment}\
  }\textbf {\bibinfo {volume} {2023}},\ \bibinfo {pages} {043202} (\bibinfo
  {year} {2023})}\BibitemShut {NoStop}%
\bibitem [{\citenamefont {Wijns}\ \emph {et~al.}(2024)\citenamefont {Wijns},
  \citenamefont {Eichhorn},\ and\ \citenamefont
  {Cleuren}}]{wijns2023microscopic}%
  \BibitemOpen
  \bibfield  {author} {\bibinfo {author} {\bibfnamefont {B.}~\bibnamefont
  {Wijns}}, \bibinfo {author} {\bibfnamefont {R.}~\bibnamefont {Eichhorn}},\
  and\ \bibinfo {author} {\bibfnamefont {B.}~\bibnamefont {Cleuren}},\
  }\bibfield  {title} {\bibinfo {title} {Microscopic model for a brownian
  translator},\ }\href {https://doi.org/10.1088/1742-5468/ad3199} {\bibfield
  {journal} {\bibinfo  {journal} {Journal of Statistical Mechanics: Theory and
  Experiment}\ }\textbf {\bibinfo {volume} {2024}},\ \bibinfo {pages} {043203}
  (\bibinfo {year} {2024})}\BibitemShut {NoStop}%
\bibitem [{\citenamefont {Reimann}(2002)}]{reimannPR2002}%
  \BibitemOpen
  \bibfield  {author} {\bibinfo {author} {\bibfnamefont {P.}~\bibnamefont
  {Reimann}},\ }\bibfield  {title} {\bibinfo {title} {Brownian motors: noisy
  transport far from equilibrium},\ }\href@noop {} {\bibfield  {journal}
  {\bibinfo  {journal} {Physics Reports}\ }\textbf {\bibinfo {volume} {361}},\
  \bibinfo {pages} {57} (\bibinfo {year} {2002})}\BibitemShut {NoStop}%
\bibitem [{\citenamefont {Klajn}\ \emph {et~al.}(2010)\citenamefont {Klajn},
  \citenamefont {Stoddart},\ and\ \citenamefont {Grzybowski}}]{KlajnCSR2010}%
  \BibitemOpen
  \bibfield  {author} {\bibinfo {author} {\bibfnamefont {R.}~\bibnamefont
  {Klajn}}, \bibinfo {author} {\bibfnamefont {J.~F.}\ \bibnamefont
  {Stoddart}},\ and\ \bibinfo {author} {\bibfnamefont {B.~A.}\ \bibnamefont
  {Grzybowski}},\ }\bibfield  {title} {\bibinfo {title} {Nanoparticles
  functionalised with reversible molecular and supramolecular switches},\
  }\href {https://doi.org/10.1039/B920377J} {\bibfield  {journal} {\bibinfo
  {journal} {Chem. Soc. Rev.}\ }\textbf {\bibinfo {volume} {39}},\ \bibinfo
  {pages} {2203} (\bibinfo {year} {2010})}\BibitemShut {NoStop}%
\bibitem [{\citenamefont {Liu}\ \emph {et~al.}(2012)\citenamefont {Liu},
  \citenamefont {Bastiaansen}, \citenamefont {den Toonder},\ and\ \citenamefont
  {Broer}}]{LiuACIE2012}%
  \BibitemOpen
  \bibfield  {author} {\bibinfo {author} {\bibfnamefont {D.}~\bibnamefont
  {Liu}}, \bibinfo {author} {\bibfnamefont {C.~W.~M.}\ \bibnamefont
  {Bastiaansen}}, \bibinfo {author} {\bibfnamefont {J.~M.~J.}\ \bibnamefont
  {den Toonder}},\ and\ \bibinfo {author} {\bibfnamefont {D.~J.}\ \bibnamefont
  {Broer}},\ }\bibfield  {title} {\bibinfo {title} {Photo-switchable surface
  topologies in chiral nematic coatings},\ }\href
  {https://doi.org/https://doi.org/10.1002/anie.201105101} {\bibfield
  {journal} {\bibinfo  {journal} {Angewandte Chemie International Edition}\
  }\textbf {\bibinfo {volume} {51}},\ \bibinfo {pages} {892} (\bibinfo {year}
  {2012})},\ \Eprint
  {https://arxiv.org/abs/https://onlinelibrary.wiley.com/doi/pdf/10.1002/anie.201105101}
  {https://onlinelibrary.wiley.com/doi/pdf/10.1002/anie.201105101} \BibitemShut
  {NoStop}%
\bibitem [{\citenamefont {Liu}\ \emph {et~al.}(2026)\citenamefont {Liu},
  \citenamefont {Nguyen}, \citenamefont {Lin}, \citenamefont {Sun},\ and\
  \citenamefont {Zheng}}]{LiuCR2026}%
  \BibitemOpen
  \bibfield  {author} {\bibinfo {author} {\bibfnamefont {S.-F.}\ \bibnamefont
  {Liu}}, \bibinfo {author} {\bibfnamefont {K.}~\bibnamefont {Nguyen}},
  \bibinfo {author} {\bibfnamefont {L.}~\bibnamefont {Lin}}, \bibinfo {author}
  {\bibfnamefont {H.-B.}\ \bibnamefont {Sun}},\ and\ \bibinfo {author}
  {\bibfnamefont {Y.}~\bibnamefont {Zheng}},\ }\bibfield  {title} {\bibinfo
  {title} {Optical colloidal assembly},\ }\href
  {https://doi.org/10.1021/acs.chemrev.5c00644} {\bibfield  {journal} {\bibinfo
   {journal} {Chemical Reviews}\ }\textbf {\bibinfo {volume} {126}},\ \bibinfo
  {pages} {448} (\bibinfo {year} {2026})},\ \bibinfo {note} {pMID: 41364543},\
  \Eprint {https://arxiv.org/abs/https://doi.org/10.1021/acs.chemrev.5c00644}
  {https://doi.org/10.1021/acs.chemrev.5c00644} \BibitemShut {NoStop}%
\bibitem [{\citenamefont {Kim}\ and\ \citenamefont
  {Karrila}(2013)}]{kim2013microhydrodynamics}%
  \BibitemOpen
  \bibfield  {author} {\bibinfo {author} {\bibfnamefont {S.}~\bibnamefont
  {Kim}}\ and\ \bibinfo {author} {\bibfnamefont {S.~J.}\ \bibnamefont
  {Karrila}},\ }\href@noop {} {\emph {\bibinfo {title} {Microhydrodynamics:
  principles and selected applications}}}\ (\bibinfo  {publisher}
  {Butterworth-Heinemann},\ \bibinfo {year} {2013})\BibitemShut {NoStop}%
\bibitem [{\citenamefont {Carrasco}\ and\ \citenamefont {Garc{\i}a de~la
  Torre}(1999)}]{carrascoJCP1999}%
  \BibitemOpen
  \bibfield  {author} {\bibinfo {author} {\bibfnamefont {B.}~\bibnamefont
  {Carrasco}}\ and\ \bibinfo {author} {\bibfnamefont {J.}~\bibnamefont
  {Garc{\i}a de~la Torre}},\ }\bibfield  {title} {\bibinfo {title} {Improved
  hydrodynamic interaction in macromolecular bead models},\ }\href@noop {}
  {\bibfield  {journal} {\bibinfo  {journal} {The Journal of Chemical Physics}\
  }\textbf {\bibinfo {volume} {111}},\ \bibinfo {pages} {4817} (\bibinfo {year}
  {1999})}\BibitemShut {NoStop}%
\bibitem [{\citenamefont {Carrasco}\ and\ \citenamefont
  {De~La~Torre}(1999)}]{carrascoBJ1999}%
  \BibitemOpen
  \bibfield  {author} {\bibinfo {author} {\bibfnamefont {B.}~\bibnamefont
  {Carrasco}}\ and\ \bibinfo {author} {\bibfnamefont {J.~G.}\ \bibnamefont
  {De~La~Torre}},\ }\bibfield  {title} {\bibinfo {title} {Hydrodynamic
  properties of rigid particles: comparison of different modeling and
  computational procedures},\ }\href@noop {} {\bibfield  {journal} {\bibinfo
  {journal} {Biophysical Journal}\ }\textbf {\bibinfo {volume} {76}},\ \bibinfo
  {pages} {3044} (\bibinfo {year} {1999})}\BibitemShut {NoStop}%
\end{thebibliography}%
\appendix
\section{Tracer Dynamics}
The dynamics of the tracer object between resets is fully determined by the random collisions with the surrounding gas particles (mass $m$). The effect of a single collision, expressed by the collision rule, is determined under the conditions that these collisions are instantaneous, elastic, and subject to the constraints that the object only performs translational motion. These conditions allow us to uniquely determine the collision rule, expressed by the change in velocity $\Delta \vec{V}$ of the object, moving with velocity $\vec{V}$ prior to the collision, due to a collision with a gas particle moving with velocity $\vec{v}$,
\begin{equation}
    \Delta \vec{V}=-\frac{2\ve^2}{1+\ve^2}\left[\left(\vec{V}-\vec{v}\right).\hat{n}\right]\hat{n}.
\end{equation}
Here we introduce $\ve^2=m/M$, and $\hat{n}=\left( \sin \theta, -\cos \theta \right)$ is the normal vector on the surface at the point of impact. As these collisions are random events, $\vecv(t)$ is a stochastic process described by a probability density $P(\vecv;t)$ which satisfies the following master equation
\begin{widetext}
\begin{equation}
    \frac{\partial}{\partial t}P(\vecv,t)=\int \rmd \Delta \vecv \left[ R(\vecv - \Delta \vecv;\Delta \vecv)P(\vecv- \Delta \vecv,t)- R(\vecv|\Delta \vecv)P(\vecv',t)\right]
\end{equation}
The transition rate $R( \vec{V};\Delta \vec{V})$ captures the effect of the collisions along the whole surface. As the velocities of the incoming gas particles are Maxwellian distributed, the expression for the transition rate is
\begin{equation}
R( \vec{V};\Delta \vec{V})=
\int \rmd S \iint \rmd \vec{v} \rho\phi(\vec{v}) 
\, \Theta\left[(\vec{V}-\vec{v})\cdot\hat{n}\right]\left|(\vec{V}-\vec{v})\cdot\hat{n}\right|
\delta\left[\Delta \vec{V}+\frac{2\ve^2}{1+\ve^2}\left[\left(\vec{V}-\vec{v}\right).\hat{n}\right]\hat{n} \right].
\end{equation}
with
\begin{equation}
    \phi(\vec{v})=\frac{m}{2\pi \kB T}e^{-\frac{m}{2\kB T}\left(v_x^2+v_y^2 \right)} \; .
\end{equation}
The expression for the transition rate explicitly assumes the object is struck by gas particles whose velocities are Maxwellian distributed. This is the case for gas particles upon their first collision with the object. Their post-collisional velocities, however, are not Maxwellian distributed. Hence, in order for the transition rates to be accurate, gas particles should collide only once and then drift off afterwards. This is accomplished by using heavy, convex objects. Heavy means the mass of the object $M$ is much larger than the mass of the gas particles $m$, so $\ve^2 \ll 1 $. In this limit, the probability for the object to overtake the gas particle is strongly reduced. Convexity ensures that the gas particle is directed away from the object after the collision.\newline
By introducing the jump moments, defined as
\begin{equation}
    A_{jk}(\vecv) \equiv \int \rmd \Delta \vecv \; \dvx^j \; \dvy^k \; R(\vecv;\Delta \vecv),
\end{equation}
the master equation is used to obtain an equivalent and infinite set of evolution equations for the velocity moments,
\begin{equation}
    \frac{d}{dt} \langle V_x^m V_y^n \rangle =\sum_{k=0}^{m}\sum_{l=0}^{n}\binom{m}{k}\binom{n}{l}\left\langle V_x^{m-k}V_y^{n-l}A_{kl}(\vecv)\right\rangle - \left\langle V_x^{m}V_y^{n}A_{00}(\vecv)\right\rangle
\, .
\end{equation}
\end{widetext}
Rescaling time and velocity,
\begin{equation}
    v_x = \sqrt{\frac{M}{kT}}V_x \;\;\;\; ; \;\;\;\; v_y = \sqrt{\frac{M}{kT}}V_y \; ,
\end{equation}
and a subsequent series expansion in $\ve$ decouples these equations, and we eventually end up with the equations given in Eq.~\eqref{eq:moments}.

A final note concerns the goniometric averages, cf Eq.~\eqref{eq:gonio}, appearing as coefficients in the equations for the moments. These averages involve the the shape function $s(\theta)$, defined such that $s(\theta)\ud \theta$
is the fraction of the surface with orientation $\theta$ w.r.t the
$x$-axis. For the convex polygons we consider in this work, the shape function becomes a weighted sum of Dirac delta functions:
\begin{equation}
    s(\theta)=\sum_i \frac{\ell_i}{S} \delta \left(\theta - \theta_i \right) \; ,
\end{equation}
with $\ell_i$ and $\theta_i$ respectively the length and angle of line segment $i$, and $S=\sum_i \ell_i$ the total circumference of the polygon.
\section{Colloidal Particles: Mobility tensor}
\label{app:mu}
In general, one would have to solve the Stokes equation to calculate
the mobility tensor $\mu$ for a given particle shape
\cite{kim2013microhydrodynamics}. However, for particles that are rigidly assembled from spherical beads,
there is a well-established procedure to calculate $\mu$ from the hydrodynamic interactions between all
component beads \cite{carrascoJCP1999} (see also \cite{carrascoBJ1999}).
We employ this procedure for the different colloidal particles shown in Fig.~\ref{fig:colloid_separation}
to calculate $\mu$ in a \textit{body-fixed reference frame},
which corresponds to $\varphi=0$.

The four different particle species we show in Fig.~\ref{fig:colloid_separation} are assembled from hard spheres of diameter $1.0$. Using a viscosity of $1.0$ for the aqueous environment, the results for $\mu$ are:
\begin{itemize}
\item Green, ``cross-shaped'' particle,
\begin{subequations}
\label{eq:mus}
\begin{equation}
\mu = \left(
\begin{array}{ccc}
0.05354 & 0.0 & 0.0 \\
0.0 & 0.05354 & 0.0 \\
0.0 & 0.0 & 0.02302  \\
\end{array}
\right)
\, .
\end{equation}
\item Yellow, ``T-shaped'' particle,
\begin{equation}
\mu = \left(
\begin{array}{ccc}
0.05165 & 0.0 & 0.0 \\
0.0 & 0.05091 & -0.001761 \\
0.0 & -0.001761 & 0.01805 \\
\end{array}
\right)
\, .
\end{equation}
\item Red, ``L-shaped'' particle,
\begin{equation}
\mu = \left(
\begin{array}{ccc}
0.05962 & 0.002094 & -0.001775 \\
0.002094 & 0.05494 & -0.002674 \\
-0.001775 & -0.002674 & 0.02501 \\
\end{array}
\right)
\, .
\end{equation}
\item Blue, ``$\Gamma$-shaped'' particle,
\begin{equation}
\mu = \left(
\begin{array}{ccc}
0.05962 & -0.002094 & 0.001775 \\
-0.002094 & 0.05494 & -0.002674 \\
0.001775 & -0.002674 & 0.02501 \\
\end{array}
\right)
\, .
\end{equation}
\end{subequations}
\end{itemize}
\end{document}